\documentclass[draft]{aipproc}
\layoutstyle{6x9}

\def\d{{\rm d}}
\def\dt{\partial_t}
\def\Dt{\Delta_t}

\def\ep{\epsilon}

\def\tr{{\rm tr}\,}

\begin{document}

\title{On the Noncommutativity Approach to Supersymmetry on the Lattice}


\author{Falk Bruckmann}
  {address={Institut f\"ur Theoretische Physik, 
            Universit\"at Regensburg,
            D-93040 Regensburg (speaker)}}

\author{Mark de Kok}
  {address={Instituut-Lorentz, 
            Universiteit Leiden,
            P.O. Box 9506,
            NL-2300 RA Leiden}}

\begin{abstract} The noncommutativity approach to SUSY on the lattice 
is shown to be inconsistent and a similar inconsistency is displayed for the link approach.
\end{abstract}

\maketitle

Supersymmetry (SUSY) is a celebrated symmetry relating bosons and fermions
which may be realized in particle physics
and has many interesting theoretical features.
It would be nice to be able to investigate SUSY numerically on the lattice.
To this end, an exact SUSY invariance of the lattice action would be very useful
(in analogy to exact lattice gauge invariance).
As will be shown below, this is deeply related to keeping
the Leibniz rule on the lattice,
which is the aim of the noncommutativity approach.

The simplest example of a SUSY field theory is the 1D 
{\bf supersymmetric quantum mechanics} (SUSYQM, see e.g.\ \cite{cooper:82}). 
In the continuum it is described by the algebra 
\begin{equation}
\{Q_1,Q_1\}=\{Q_2,Q_2\}=2H, \quad \{Q_1,Q_2\}=0
\end{equation}
where $H=\dt$ is the (only) generator of the (Euclidean) Poincar\'e algebra.
The simplest multiplet consists of a real boson $\phi$, 
two Majorana fermions $\psi_{1,2}$ and an auxiliary boson $D$.
The action is taken to be
\begin{eqnarray}
S=\int \d t [
\frac{1}{2}(\partial_t\phi)^2 - \frac{1}{2}D^2 
-\frac{1}{2}(\psi_1\partial_t\psi_1 +\psi_2\partial_t\psi_2)
-i(m+3g\phi^2) \psi_1\psi_2 - D(m\phi+g\phi^3)]
\label{eqn_S_explicit}
\end{eqnarray}
Upon integrating out the nondynamical field $D$, 
mass and interaction terms for $\phi$ are generated.
This action is invariant, 
$S(\phi,\psi_1,\ldots)=S(\phi+\delta_i\phi,\psi_1+\delta_i\psi_1,\ldots)$, 
under the following variations $\delta_i=\ep^i Q_i,\:i=1,2$ (no sum, $\ep_i$ fermionic): 

\begin{table}[h!]
\begin{tabular}{c|cccc}
 $\Phi$ & $\phi$ & $\psi_1$ & $\psi_2$ & $D$\\
  \hline
  $\delta_1\Phi$ & $i\epsilon^1\psi_1$ & $i\epsilon^1\partial_t\phi $ &
      $\epsilon^1 D$ & $-\epsilon^1\partial_t\psi_2$ \\
  $\delta_2\Phi$ & $i\epsilon^2\psi_2$ & $-\epsilon^2 D$ & 
      $i\epsilon^2\partial_t\phi $ &  $\epsilon^2\partial_t \psi_1$
 \end{tabular} 
\label{table_SQMtrans}
\end{table}

\noindent Several aspects become clearer in the superfield formalism 
where the fields are organized into components of a Hermitian superfield
\begin{equation}
\Phi(t,\theta^1,\theta^2) =
\phi(t) +
i\theta^1 \psi_1(t) +
i\theta^2 \psi_2(t) +
i\theta^2\theta^1 D(t)
\end{equation}
The $\theta$'s are Grassmann coordinates
and the supercharges can be represented as 
$Q_i = \partial_{\theta^i} + \theta^i \partial_t$  
in close analogy to $H=\dt$.
The action is constructed as
\begin{equation}
S = S_{\rm kin} + S_{\rm pot},\quad
S_{\rm kin} = \int \d t\: \d^2\theta\: \frac{1}{2}D_2\Phi D_1\Phi,\quad
S_{\rm pot} = \int \d t\: \d^2\theta\:i F(\Phi),
\end{equation}
with $D_i$ the superderivatives. 
Choosing $F(\Phi)=\frac{1}{2}m\Phi^2 + \frac{1}{4}g\Phi^4$
and integrating out the $\theta$'s gives the action in the form of Eq.\ (\ref{eqn_S_explicit}).

Why $S$ is invariant will be demonstrated now for the kinetic term 
(skipping indices $i$; the potential term is invariant for the same reasons):
\begin{eqnarray}
\delta S_{\rm kin}\equiv S_{\rm kin}[\Phi+\delta\Phi]-S_{\rm kin}[\Phi]
\!\!\!\!&=&\!\!\!\!\!\!\int \d t\: \d^2\theta\:\frac{1}{2} [D_2(\ep Q\Phi)D_1\Phi+
D_2\Phi D_1(\ep Q \Phi)]\\
\!\!\!\!&=&\!\!\!\!\!\!\int \d t\: \d^2\theta\:\frac{1}{2} [(\ep Q)D_2\Phi\cdot D_1\Phi+
D_2\Phi \cdot(\ep Q) D_1\Phi]\nonumber
\end{eqnarray}
We used that the $D_i$ anticommute with $Q_j$ (and $\ep^j$).
The crucial step is the following:
\begin{eqnarray}
\delta S_{\rm kin}=\int \d t\: \d^2\theta\:\frac{1}{2}(\ep Q)[D_2\Phi D_1\Phi]=0,\qquad
\ep Q=\ep\partial_\theta+\ep\theta\partial_t
\end{eqnarray}
where we used {\bf the Leibniz rule of} $\mathbf{\ep Q}$ as a derivative operator.
$S$ is invariant because of the total derivatives in $t$ and $\theta$ as integrands
(just like for time translations).

The {\bf problem on the lattice} comes from the discretizing 
the derivative $\dt$ to $\Dt$,
the forward difference (here defined over two lattice spacings $a$; 
the backward difference works in an analogous way). This operator fulfills
\begin{equation}
\Dt [f(t)g(t)]
= [\Dt f(t)]g(t) +
f(t+ 2a)[\Dt g(t)]
\label{eqn_mod_L}
\end{equation}
Hence the Leibniz rule is violated on the lattice. 
Consequently, the naive application of continuum methods fails 
and the question is how to write down lattice actions, 
especially interacting theories, 
with exact SUSY invariance.

Most alternative attempts (see \cite{giedt:06} for a review) have part of the SUSY implemented 
exactly at finite lattice spacing and hope (sometimes prove) 
to keep typical SUSY phenomena 
without fine-tuning problems in the continuum limit.

Now we come to the {\bf noncommutativity approach} suggested by D'Adda et al. \cite{dadda:04a}.
Its idea is to have an ordinary Leibniz rule for $\delta=\ep Q$ by turning 
the modified Leibniz rule (\ref{eqn_mod_L}) of the lattice difference $\Dt$
into an ordinary Leibniz rule for $\ep\theta\Dt$:
\begin{equation}
\ep\theta\Dt [f(t)g(t)]
= [\ep\theta\Dt f(t)]g(t) +
f(t)[\ep\theta\Dt g(t)]
\end{equation} 
and keeping this rule for $\ep\partial_\theta$. This can be done through the noncommutativities
\begin{equation}
[t,\theta^{i}] = a\theta^i,\quad
[t,\partial_{\theta^i}] = -a\partial_{\theta^i},\quad
[t,\epsilon^{i}] = a\epsilon^i
\label{eqn_nc}
\end{equation}

As a consequence of the nc approach specific shifts appear in the action 
(by bringing $\theta$'s to the left to be integrated out). 
For example the mass terms in SUSYQM read \cite{bruckmann:06a}:
\begin{equation}
S_m=a\sum_t-m[i\psi_1(t+a)\psi_2(t)-i\psi_2(t+a)\psi_1(t)+D(t)\{\phi(t)+\phi(t+2a)\}/2]
\label{eqn_action_shifted}
\end{equation}
This action is supposed to be invariant under {\em all} variations $\delta_i=\ep^i Q_i$ 
(see the table on the first page)
when using the noncommutativity of $t$ and $\ep^i$, Eq.\ (\ref{eqn_nc}), 
in products of fields.
The authors of \cite{dadda:04a} claim the nc approach works for  
theories with $N=D=2$ and $N=D=4$, whereas above we have applied it to the simplest case, 
SUSYQM \cite{bruckmann:06a}.


However, there is an {\bf inconsistency} in the nc approach: 
the SUSY variations of a product of fields depends on their order \cite{bruckmann:06a}.
On the one hand, two fields are varied as
\begin{eqnarray}
fg\to (f+\ep Q f)(g+\ep Q g)=fg+\ep(Q f(t)\cdot g(t)+f(t+a)\cdot Q g(t))
\label{eqn_var_first}
\end{eqnarray}
Interchanging the order (for simplicity restricting to bosonic fields $f$ and $g$) gives
\begin{eqnarray}
gf\to (g+\ep Q g)(f+\ep Q f)
&\!\!\!\!=&\!\!\!\!gf+\ep(Q g(t)\cdot f(t)+g(t+a)\cdot Q f(t))\nonumber\\
&\!\!\!\!=&\!\!\!\!fg+\ep(Q f(t)\cdot g(t+a)+f(t)\cdot Q g(t))
\label{eqn_var_second}
\end{eqnarray}
These variations, Eq.s (\ref{eqn_var_first}) and (\ref{eqn_var_second}), 
do not agree due to the different shifts, 
but they have to as they come from the product of two commuting fields!

As a consequence, when checking the invariance of the action, the expression
$D(t)\phi(t)-i\psi_2(t+a)\psi_1(t)$ (see (\ref{eqn_action_shifted}))
gives a total derivative under the variations $\ep^1 Q_1$ 
(just like in the continuum), but $\phi(t)D(t)-i\psi_2(t+a)\psi_1(t)$ 
does not: because of the different shifts cancellations have been destroyed.
Hence the two equivalent forms of this term give 
different answers concerning the SUSY invariance of it!

In \cite{bruckmann:06a} we have shown this inconsistency to
be a generic feature of the noncommutativity approach.
The rationale of this inconsistency lies in the fact
that the noncommutativity $[t,\ep]\neq 0$ forbids to treat $t$ as a number.
Correspondingly, the component fields cannot be ordinary functions 
and this theory cannot be simulated numerically.

We would like to add a related finding. The {\bf link approach} to lattice SUSY, 
emanating from the nc approach and proposed by the same authors 
in \cite{dadda:05a}, suffers from {\bf a similar inconsistency}. 
In \cite{dadda:05a}
the SUSY transformations $s_A$ are defined 
via the (anti)commutator with a fermionic link $\nabla_A$. 
This link nature gives $s_A$ a modified Leibniz rule, 
much like in Eq.\ (\ref{eqn_var_first}). 
For example, the trace over two bosonic site variables $f$ and $g$ transforms as 
\begin{equation}
s_A[\tr f_{x,x}g_{x,x}]=\tr (s_A f)_{x+a_A,x}g_{x,x}+\tr f_{x+a_A,x+a_A}(s_A g)_{x+a_A,x}
\end{equation}
In order to arrive at the analogue of Eq. (\ref{eqn_var_second}) we consider the inverse order
\begin{equation}
s_A[\tr g_{x,x}f_{x,x}]=\tr (s_A g)_{x+a_A,x}f_{x,x}+\tr g_{x+a_A,x+a_A}(s_A f)_{x+a_A,x}
\end{equation}
Because of the traces, the (matrix-valued) fields on the l.h.s.s can be interchanged. 
But the results on the r.h.s.s differ by shifts again!
As will be shown in a forthcoming publication \cite{bruckmann:06b}, 
this ambiguity also plaques the action of the link approach, 
which therefore is inconsistent, too. \\ 

FB likes to thank the organizers for a very nice conference. We are grateful to 
Simon Catterall, Joel Giedt, David Kaplan and the Jena group for helpful discussions. 


\end{document}